\documentclass[%
 reprint,
superscriptaddress,
 amsmath,amssymb,longbibliography,
 aps,
prr,
]{revtex4-1}
\usepackage{hyperref}

\usepackage{graphicx}
\usepackage{amsmath}
\usepackage{rotating}
\usepackage{color}

\newcommand{\sruo}{Sr$_2$RuO$_4$}
\newcommand{\lsoc}{\lambda_{so}}
\newcommand{\io}{\tilde \mu}

\newcommand{\down}{\downarrow}
\newcommand{\up}{\uparrow}
\newcommand{\spin}{\sigma}

\newcommand{\kv}{{\bf k}}
\newcommand{\qv}{{\bf q}}
\newcommand{\Qv}{{\bf Q}}

 \usepackage{tabularx}
 
\begin{document}

\title{Leading superconducting instabilities in three-dimensional models for Sr$_2$RuO$_4$}

\author{Astrid T. R{\o}mer}
\affiliation{Niels Bohr Institute, University of Copenhagen, 2100 Copenhagen, Denmark}
\author{T. A. Maier}
\affiliation{Center for Nanophase Materials Sciences, Oak Ridge National Laboratory, Oak Ridge, Tennessee 37831, USA}
\author{Andreas Kreisel}
\affiliation{Institut f\" ur Theoretische Physik, Universit\"at Leipzig, D-04103 Leipzig, Germany}
\author{P. J. Hirschfeld}
\affiliation{Department of Physics, University of Florida, Gainesville, Florida 32611, USA}
\author{Brian M. Andersen}
\affiliation{Niels Bohr Institute, University of Copenhagen, 2100 Copenhagen, Denmark}

\date{\today}
\begin{abstract}
The unconventional superconductor \sruo~has been the subject of enormous interest over more than two decades, but until now the form of its order parameter has not been explicitly determined. Since groundbreaking NMR experiments revealed recently that the pairs are of dominant spin-singlet character, attention has focused on time-reversal symmetry breaking linear combinations of $s$-, $d$- and $g$-wave one-dimensional (1D) irreducible representations. However, a state of the form $d_{xz}+id_{yz}$ corresponding to the two-dimensional representation $E_g$ has also been proposed based on some experiments. We present a systematic study of the stability of various superconducting candidate states, assuming that pairing is driven by the fluctuation exchange mechanism, including  a realistic three-dimensional Fermi surface, full treatment of both local and non-local spin-orbit couplings, and a wide range of Hubbard-Kanamori interaction parameters $U,J,U',J'$. The leading superconducting instabilities are found to exhibit nodal even-parity $A_{1g} (s')$ or $B_{1g} (d_{x^2-y^2})$ symmetries, similar to the findings in two-dimensional models without longer-range Coulomb interaction which tends to favor $d_{xy}$ over $d_{x^2-y^2}$. Within the so-called Hund's coupling mean-field pairing scenario, the $E_g (d_{xz}/d_{yz})$ solution can be stabilized for large $J$ and specific forms of the spin-orbit coupling, but for all cases studied here the eigenvalues of other superconducting solutions are significantly larger when the full fluctuation exchange vertex is included in the pairing kernel. Additionally, we compute the spin susceptibility in relevant superconducting candidate phases and compare to recent neutron scattering and NMR Knight shift measurements. It is found that $d_{xz}+id_{yz}$ order supports a neutron resonance in its superconducting phase, in contrast to a recent experiment [K. Jenni {\it et al.}, Phys. Rev. B {\bf 103}, 104511 (2021)], whereas  $s'+id_{x^2-y^2}$ does not. Furthermore, comparison of the Knight shift reveals that $s'+id_{x^2-y^2}$ exhibits a larger low-temperature shift than $d_{xz}+id_{yz}$.

\end{abstract}

\maketitle

\section{Introduction} 

The superconducting state of \sruo~remains unresolved and
after nearly three decades of study, the current lack of consensus on the nature of the superconducting ground state highlights the need for better  quantitative theoretical descriptions of unconventional superconductivity~\cite{Mackenzie2017}.  There is  consensus on some qualitative aspects: for example, recent ultrasound experiments impose important constraints on the possibilities of superconducting orders in \sruo, suggesting composite pairing solutions where the product of two distinct superconducting components transforms as $B_{2g}$ ($d_{xy}$)~\cite{ghosh2020thermodynamic,benhabib2020jump}.
In two-dimensional (2D) models, this leaves open the possibility of nearly degenerate superconducting instabilities belonging to different irreducible representations forming solutions of the form $d_{x^2-y^2}+ig_{(x^2-y^2)xy}$~\cite{kivelson2020proposal,willa2021} or $s'+id_{xy}$~\cite{clepkens2021,romer2021}. (We denote by $s'$ ($s$) order parameters with nodal (nodeless) $A_{1g}$ symmetry.) The 2D representation $E_u:p_x/p_y$ seems ruled out due to a pronounced drop in Knight shift observed by nuclear magnetic resonance (NMR)~\cite{Pustogow19,Ishida_correct,Chronister2021}. However, as pointed out recently, when considering the third spatial dimension there exists another 2D irreducible representation of the $D_{4h}$ point group which could potentially encompass the observations from NMR, ultrasound experiments and $\mu$SR, namely the even-parity solution $E_g:d_{xz}/d_{yz}$~\cite{Pustogow19,ghosh2020thermodynamic,Luke1998,Grinenko2021}. This state, however, exhibits a controversial nodal structure with (horizontal) line nodes at $k_z=0$~\cite{Hassinger17}, and relies on interlayer Cooper pairing despite the pronounced 2D electronic structure of \sruo.

\begin{figure*}[t!]
    \centering
    \includegraphics[width=0.9\linewidth]{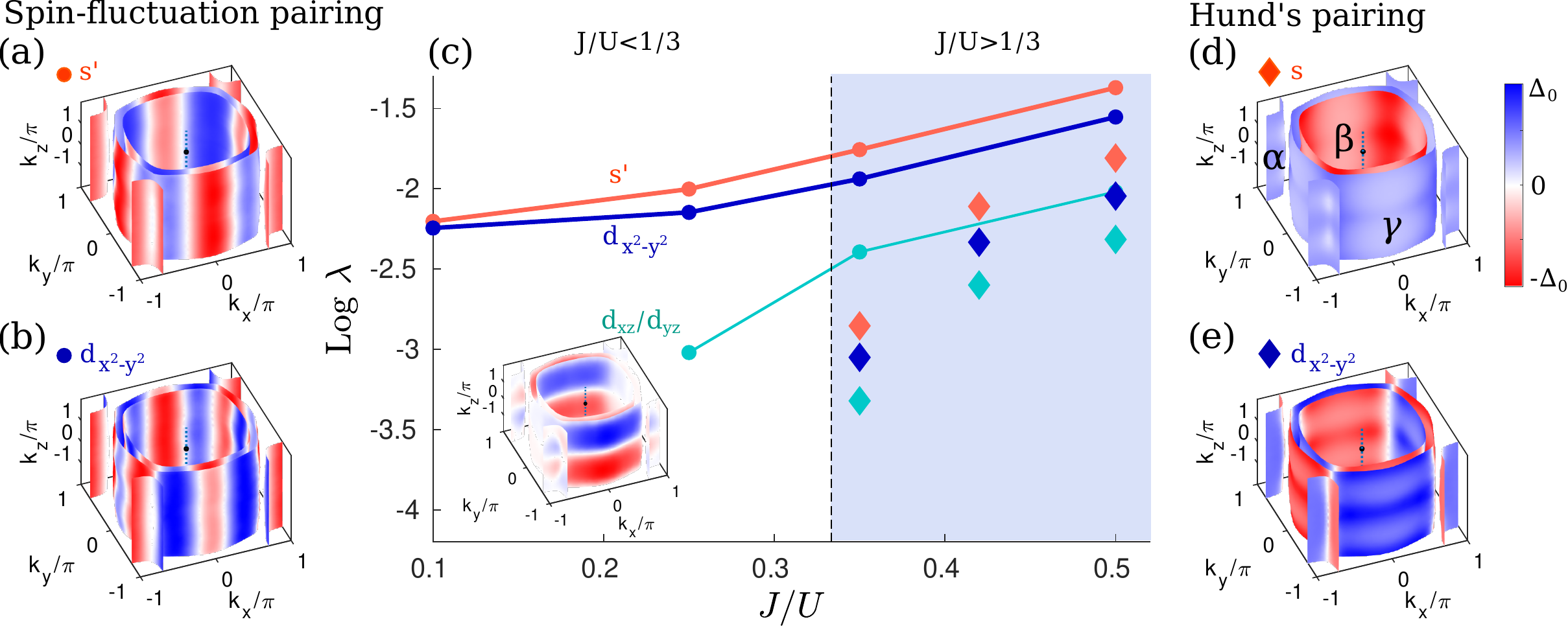}
    \caption{(a,b) Leading nodal $s'$ and subleading  $d_{x^2-y^2}$ gap structures arising within spin-fluctuation mediated pairing for $U=100$ meV and $J/U=0.25$. (c) Evolution of the eigenvalues of the linearized gap equation [Eq.~(\ref{eq:LGE})] as a function of Hund's coupling for onsite Coulomb repulsion $U=100$ meV. Note the logarithmic scale. The orange (blue) line displays the eigenvalue of the $s'$ ($d_{x^2-y^2}$) solutions, while the turquoise line shows the eigenvalue of the $d_{xz}/d_{yz}$ solution, all obtained within the spin-fluctuation mediated pairing scenario. Diamond symbols display the eigenvalues of the local Hund's coupling pairing scenario with $s$ shown by orange, $d_{x^2-y^2}$ shown by blue diamonds and $d_{xz}/d_{yz}$ displayed with turquoise diamonds. The lower inset shows one of the two components of the $E_g$ ($d_{xz}/d_{yz}$) solution. (d,e) Leading gap structures of the Hund's coupling pairing scenario with $J/U=0.5$. Note that these Hund's pair states correspond to dominant inter-orbital singlet, spin-triplet pairs~\cite{vafek,clepkens_PRB_2021}. In (d) the usual convention for denoting the Fermi surface pockets is indicated. The colorbar shown in this plot is common for all gap plots.}
    \label{fig:Cobo1}
\end{figure*}

Most previous theoretical works addressing the gap structure of \sruo~were restricted to the two spatial dimensions of the RuO$_2$ planes. This approximation appears natural based on the pronounced 2D structure of the system as observed by e.g. the resistivity anisotropy ratio~\cite{maeno94}, de Hass-van Alphen oscillations~\cite{Bergemann_2003}, and angular resolved photoemission spectroscopy (ARPES)~\cite{Damascelli_2000,Haverkort_2008,Zabolotnyy13}. However, the fact that the orbital content at the Fermi surface varies along $k_z$  as a result of spin-orbit coupling (SOC)~\cite{Veenstra2014} makes this state more plausible. Recently, the importance of the third dimension on the nature of the superconducting state was explored in
Refs.~\onlinecite{fukaya2021spin,clepkens2021,suh2019,Roising2019,Huang18,Jerzembeck_2021}. In particular, local pairing originating from large Hund's couplings treated within mean-field theory can stabilize spin-triplet, orbital-singlet superconductivity\cite{vafek,Puetter_2012}, which despite the spin-triplet structure can be compatible with a Knight shift reduction~\cite{suh2019,Yu_2018,fukaya2021spin,Gupta_Annett_2021}. 

When onsite attractive interactions arise from large Hund's coupling, the superconducting gap structure is inherited from the normal state band structure. This gives rise to an  inter-orbital spin-triplet pairing gap if onsite SOC dominates~\cite{vafek,Cheung_2019}. However, other solutions can materialize in the presence of sizable  momentum-dependent spin-orbit couplings (k-SOC). For example, the $E_g$ solution can be stabilized when the k-SOC terms exhibit $\sin k_x\sin k_z$ or $\sin k_y\sin k_z$ structure, as was demonstrated recently in the case of \sruo~\cite{suh2019,clepkens2021,clepkens_PRB_2021}. In contrast to the "Hund's coupling pairing scenario" based on constant attractive interactions, spin-fluctuation mediated pairing provides a route to superconductivity irrespective of the strength of SOC and Hund's coupling. In this mechanism, attractive channels are promoted by nesting features of the Fermi surface pockets, enabling a number of near-degenerate superconducting instabilities. These solutions belong to different irreducible representations of the $D_{4h}$ point group and have highlighted the concept of accidental degeneracies in the case of \sruo~\cite{Eschrig2001,Scaffidi2014,Wang19,WangKallin20,RomerPRL,rmer2020fluctuationdriven,romer2021}. 

The above developments call for a closer comparison of the Hund's coupling versus spin-fluctuation-mediated pairing mechanisms for realistic 3D models for \sruo. In Ref.~\onlinecite{Merce} we have performed a direct comparison between these two mechanisms for different generic 2D band structures. Here for \sruo, we are concerned with the question of how the leading gap solutions compare in terms of their momentum structure and critical temperature $T_{\mathrm c}$. In addition, in the Hund's coupling pairing scenario, k-SOC terms are important for the resulting gap structure~\cite{suh2019,clepkens_PRB_2021}, but how do such k-SOC terms alter the spin-fluctuation generated gaps? In fact, the close competition between different pairing symmetries in \sruo~\cite{Scaffidi2014,RomerPRL,Gingras18,romer2021} makes it important to determine if such rather small changes in the underlying electronic structure can rearrange the hierarchy of superconducting solutions. Therefore, regardless of the smallness of additional interplanar hybridization and k-SOC terms that are allowed in 3D models, such terms might affect the superconducting gap symmetry. 

In this paper, we determine the leading superconducting instabilities from a 3D realistic band structure relevant for \sruo. Our calculations reveal the effect of Hund's coupling on the leading gap structures in a broad parameter range and thereby connect the pairing outcomes from the spin-fluctuation mediated mechanism with the regime where onsite Hund's coupling enables superconducting pairing, which is $J/U>\frac{1}{3}$ at the mean-field level. Figure~\ref{fig:Cobo1} provides an overview of the resulting evolution of the dominant instabilities as a function of $J/U$. The two different mechanisms agree on the overall symmetry of the gap solutions, favoring $A_{1g} (s')$ and $B_{1g} (d_{x^2-y^2})$ superconductivity. However, as seen from Fig.~\ref{fig:Cobo1}, the nodal structures of the superconducting gaps are very different, with additional nesting-enforced vertical line nodes present when superconductivity is generated by spin fluctuations, see Fig.~\ref{fig:Cobo1}(a,b). Furthermore, a comparison of the leading eigenvalues of the two different mechanisms displayed in Fig.~\ref{fig:Cobo1}(c) highlights the fact that the eigenvalues, and hence $T_{\mathrm c}$, for the Hund's pairing mechanism are much smaller compared to the spin-fluctuation mechanism.  This appears to indicate that this version of Hund's coupling pairing may not be directly  applicable to a correlated material like \sruo, which exhibits significant momentum structure in its magnetic susceptibility. Finally, we remark that longer-range Coulomb repulsion  promote $d_{xy}$ pairing, as shown recently within 2D models~\cite{romer2021}. Therefore, due to the relatively weak 3D effects in \sruo, we expect that many of the results shown below remain valid when including longer-range Coulomb repulsion except $d_{x^2-y^2}$ gets replaced by $d_{xy}$, but we have not further explored such effects in the present paper.

Additionally, we compute the dynamical spin susceptibility in relevant superconducting states for 3D \sruo. It is found that only $d_{x^2-y^2}$ and $d_{xz}/d_{yz}$ superconductivity support neutron resonances, whereas for example $s'+i d_{x^2-y^2}$ does not. Lastly, a calculation of the uniform susceptibility shows that both $s'+id_{x^2-y^2}$ and $d_{xz}+id_{yz}$ exhibit pronounced Knight shifts, even though the former structure appears more compatible with recent bounds on the residual susceptibility at the lowest temperatures~\cite{Chronister2021}. We discuss these findings in light of known experimental measurements of these quantities.  

\section{Model and method}

The presence of inversion and time-reversal symmetries in the normal state Hamiltonian allows for 15 distinct non-zero terms within the t$_{2g}$ orbital subset~\cite{Ramires2019,suh2019}, which include spin-preserving hopping terms, hybridizations, as well as onsite and momentum-dependent SOC. The latter depends explicitly on spin and includes both spin-flipping and spin-preserving processes. The full normal state Hamiltonian takes the form of a $6\times 6$ matrix
\begin{eqnarray}
    H_{\rm NS}&=&\left( \begin{array}{c  |c}
     & \\
    H_0 + H_{SOC}^{+} & H_{hyb}+H_{SOC}^{3D} \\
    & \\
    \hline
        & \\
    (H_{hyb}+H_{SOC}^{3D})^\dagger &    H_0 + H_{SOC}^{-} \\
     & \\
    \end{array}\right),
    \label{eq:H}
\end{eqnarray}
in the pseudospin basis $[\Psi(\kv,+),\Psi(\kv,-)]$. Here, $\Psi(\kv,\sigma)=[d_{\kv,xz,\sigma},d_{\kv,yz,\sigma},d_{\kv,xy,\overline\sigma}]$
and $d_{\kv,\mu, \sigma}$ denotes the annihilation of an electron in orbital $\mu$ and spin state $\sigma$ with momentum $\kv$.

In Eq.~(\ref{eq:H}) the electronic dispersion of the three orbitals is given by
\begin{eqnarray}
 H_0(\kv)&=&\left( \begin{array}{ccc}
  \xi_{xz}(\kv) & g(\kv) &0 \\
  g(\kv) & \xi_{yz}(\kv) & 0 \\
  0 & 0 &  \xi_{xy}(\kv)
  \end{array}\right),
  \label{eq:H0}
\end{eqnarray}
with $\xi_{xz/yz}(\kv)=-2t_{1/2}\cos k_x -2t_{2/1}\cos k_y -\mu$, $\xi_{xy}(\kv)=-2t_3(\cos k_x +\cos k_y) -4t_4\cos k_x \cos k_y-2t_5(\cos 2k_x +\cos 2k_y) -\mu$, and a hybridization of the form $g(\kv)=-4 t' \sin(k_x)\sin(k_y)-8 t'' \sin(k_x/2) \sin(k_y/2) \cos(k_z/2)$.
The onsite and momentum-dependent SOC terms entering in the block diagonals of Eq.~(\ref{eq:H}) are given by
\begin{eqnarray}
  H_{SOC}^{ \sigma}(\kv)&=&\frac{1}{2}\left( \begin{array}{ccc}
  0 & -i\sigma\lsoc & i\lsoc\\
  i\sigma\lsoc & 0 & -\sigma\lsoc \\
  -i\lsoc&-\sigma\lsoc & 0
\end{array}\right)\\
&+&\left( \begin{array}{ccc}
  0 & 0 & \sigma \alpha_\kv -i\beta_\kv\\
 0 & 0 &  -i \alpha_\kv -\sigma\beta_\kv \\
  \sigma \alpha_\kv +i\beta_\kv &i \alpha_\kv -\sigma\beta_\kv & 0
\end{array}\right), \nonumber
\label{eq:Hsoc}
\end{eqnarray}
where 
$\lsoc$ denotes the atomic SOC ($\bf L \cdot S$) and $\alpha_\kv,\beta_\kv$ are momentum-dependent, spin-flip SOC terms given by
\begin{eqnarray}
\alpha_\kv&=&4 t_{dxy}\sin(k_x)\sin(k_y),\\
\beta_\kv&=&2 t_d (\cos(k_x)-\cos(k_y)).
\end{eqnarray}
These higher order coupling terms can arise from the presence of orbital hybridizations and atomic SOC~\cite{clepkens2021}. Additional momentum-dependent SOC is given by the off-diagonal contributions to the normal state Hamiltonian
 \begin{eqnarray}
 H_{SOC}^{3D}(\kv)&=&\left( \begin{array}{ccc}
  0 & \eta_\kv^x-i\eta_\kv^y & -i\gamma_\kv^y \\ 
 -\eta_\kv^x+i\eta_\kv^y & 0& -i\gamma_\kv^x \\
  -i\gamma_\kv^y & -i\gamma_\kv^x& 0
\end{array}\right),
\end{eqnarray}
with
 \begin{eqnarray}
   \eta_\kv^x&=&~8t_{12z}f_x(\kv),~
\eta_\kv^y=-8t_{12z}f_y(\kv), \\
\gamma_\kv^x&=&~8 t_{56z}f_x(\kv),~
\gamma_\kv^y=-8t_{56z}f_y(\kv),
\end{eqnarray}
and form factors given by
\begin{eqnarray}
 f_x(\kv)&=& \cos(k_x/2)\sin(k_y/2)\sin(k_z/2), \label{eq:fx}\\
 f_y(\kv)&=& \sin(k_x/2)\cos(k_y/2)\sin(k_z/2). \label{eq:fy}
 \end{eqnarray}
The $\eta$-terms are spin-flip processes between the $xz$ and $yz$ orbitals, while the $\gamma$-terms quantify spin-preserving transitions between the $xz/yz$ and $xy$ orbitals~\cite{clepkens2021,suh2019}. It was previously shown that the $\gamma$-terms can give rise to an $E_g$ gap solution in the presence of local attractions arising from large Hund's couplings, as discussed in Refs.~\onlinecite{suh2019,clepkens2021}. In addition, the $\alpha_\kv$-terms of Eq.~(\ref{eq:Hsoc}) can promote an $s+id_{xy}$ solution~\cite{clepkens2021}. Since one of the goals of this study is to determine the viability of the $E_g$ state, we will focus on the $E_g$-favorable k-SOC terms with momentum structure as detailed in Eqs.~(\ref{eq:fx}-\ref{eq:fy}), i.e. finite values of $t_{12z}$ and $t_{56z}$, and set $t_{dxy}=t_d=0$.

Finally, in a 3D model, interplanar hybridizations between the $xz/yz$ and $xy$ orbitals are also allowed and enter the off-diagonal parts of Eq.~(\ref{eq:H}). These additional terms are given by
\begin{eqnarray}
  H_{hyb}(\kv)&=&\left( \begin{array}{ccc}
  0& 0& T_{xz,xy}(\kv) \\
  0 & 0& T_{yz,xy}(\kv)\\
  T_{xz,xy}(\kv) &T_{yz,xy}(\kv)&0
\end{array}\right),
\end{eqnarray}
with
\begin{eqnarray}
  T_{xz,xy}(\kv)&=&-8 t_\textrm{int} f_x(\kv),~ T_{yz,xy}(\kv)=-8 t_\textrm{int} f_y(\kv).\nonumber\\
\end{eqnarray}

\begin{table}[tb]
 \caption{Hopping  parametrization with additional hybridizations in units of meV.}
    \centering
    \begin{tabular}{|c|c|c|c|c|c|c|c|c|}
    \hline
        $t_1$ & $t_2$ & $t_3$ & $t_4$ & $t_5$ & $\mu$ & $t'$& $t''$ & $t_\textrm{int}$ \\
         \hline \hline
         $88$ & $9$ & $80$ & $40$ & $5$ &
         $109$& $3$ & $-2.5$ & $-2$ \\
         \hline
    \end{tabular}
    \label{tab:Cobohop}
\end{table}

With the above expressions for hybridization and k-SOC, 
we work in a Brillouin zone (BZ) given by  
$[-\pi,\pi]\times[-\pi,\pi]\times[-2\pi,2\pi]$. We have introduced out-of-plane hybridization terms, $t''$  and $t_\text{int}$ of similar amplitude as in Ref.~\onlinecite{suh2019} renormalized to the effective hopping parameters presented in Table~\ref{tab:Cobohop}. We fix the onsite SOC to $\lsoc=35$ meV ($\simeq 0.4 t_1$)\cite{RomerPRL} giving rise to a splitting of the bands along the zone diagonals in agreement with ARPES experiments~\cite{Zabolotnyy13}. The dependence of the pairing structure on the amplitude of $\lsoc$ has been explored in detail for 2D models in Ref.~\cite{RomerPRL}. Also, we include finite values of hybridizations (see Table ~\ref{tab:Cobohop}) as well as k-SOC parametrized by $t_{12z}=5$ meV and $t_{56z} =3,4$ meV. The interplanar k-SOC terms are chosen to be the maximum allowed values that provide a Fermi surface compatible with ARPES and introduce a significant three-dimensional spin structure (see Fig.~\ref{fig:chiRPA}). Setting all 3D couplings to zero, the band structure reduces to the 2D model investigated in Refs.~\onlinecite{RomerPRL,rmer2020fluctuationdriven,romer2021}.

The bare multi-orbital interaction Hamiltonian is given by
\begin{eqnarray}
 \hat H_{int}&=&U\sum_{i,\mu} n_{i\mu\up}n_{i\mu\down}+\frac{U'}{2}\sum_{i,\nu\neq\mu,\spin} n_{i\mu \spin}n_{i\nu\overline{\spin}}\nonumber \\
 &&+\frac{U'-J}{2}\sum_{i,\nu\neq\mu,\spin} n_{i\mu \spin}n_{i\nu \spin}\nonumber \\ &&+\frac{J}{2}\sum_{i,\nu\neq\mu,\spin} d_{i\mu \spin}^\dagger d_{i\nu\overline{\spin}}^\dagger d_{i\mu\overline{\spin}} d_{i\nu \spin}
\nonumber \\
&& +\frac{J'}{2}\sum_{i,\nu\neq\mu,\spin} d_{i\mu \spin}^\dagger d_{i\mu\overline{\spin}}^\dagger d_{i\nu\overline{\spin}} d_{i\nu \spin}, 
\label{eq:bareHint}
\end{eqnarray}
where $i$ is a site index, $\mu,\nu$ are orbital indices and $\spin=-\overline{\spin}$ refers to electronic spin. As usual, intra- and interorbital Coulomb scattering as well as Hund's coupling and pair-hopping terms are included, and $U^\prime=U-2J$, $J^\prime=J$.~\cite{Dagotto2001}
We rewrite the interaction Hamiltonian in the compact form 
\begin{align}
\hat H_{int}\!=\!\frac{1}{2}\!\!\!\sum_{ \kv,\kv' \{\tilde \mu\}}\!\!\!\Big[V(\kv,\kv')\Big]^{\io_1 , \io_2 }_{\io_3,\io_4 }d_{\kv \io_1 }^\dagger  d_{-\kv \io_3 }^\dagger d_{-\kv' \io_2 } d_{\kv' \io_4 }\nonumber \\
\label{eq:Hcooper}
\end{align}
with $\io=(\mu,\spin)$ introduced as a joint index of orbital and electronic spin. 
The bare electron-electron interaction matrix is specified by 
\begin{eqnarray}
 &&\Big[U\Big]^{\mu \spin \mu  \overline \spin}_{\mu  \overline \spin \mu \spin}=U, \qquad \Big[U\Big]^{\nu \spin \mu  \overline \spin}_{\mu  \overline \spin\nu  \spin}=U', \qquad \Big[U\Big]^{\mu \spin\nu \overline \spin}_{\mu \overline \spin\nu \spin}=J', \nonumber \\
 &&
 \Big[U\Big]^{\mu \spin\mu \overline \spin}_{\nu \overline \spin\nu \spin}=J, \qquad \Big[U\Big]^{\mu \spin\nu \spin}_{\nu \spin\mu \spin}=U'-J,
 \label{eq:Umat}
 \end{eqnarray}
 where $\mu$ and $\nu$ denote different orbitals.
 
 Within the usual mean-field Hund's coupling pairing scheme, where onsite interactions mediate superconductivity, the effective interaction of Eq.~(\ref{eq:Hcooper}) is given simply by the onsite interaction elements of Eq.~(\ref{eq:Umat}). The attractive channel obtained for $J>U'$ generates  spin-triplet, orbital-singlet superconducting order. In order to obtain a superconducting instability, however, a sizable SOC is required, since this will mix the off-Fermi level gap of spin-triplet orbital-singlet form with an intraband spin-singlet gap at the Fermi surface~\cite{Puetter_2012,vafek,Hoshino_2015,Cheung_2019,Merce}. 
 Here, we explore a range of different Hund's couplings $J/U=0.1-0.5$, noting however that in the case of \sruo~the estimated value from constrained random phase approximation (RPA) calculations is $J/U\simeq 0.1-0.2$~\cite{Mravlje11,Vaugier12,Tamai2019}.
 
In the spin-fluctuation mediated pairing mechanism, the effective interaction is derived from RPA diagrams in the normal state including onsite as well as momentum-dependent SOC. The complete pairing obtained in this framework is given by an expression resembling a multi-orbital RPA susceptibility in which the bare interactions have the multi-orbital matrix structure detailed in Eq.~(\ref{eq:Umat})~\cite{Romer2015,RomerPRL,RomerPRR2020}. The pairing kernel takes the well-known form~\cite{RomerPRL}
\begin{eqnarray}
\Big[V(\kv,\kv')\Big]^{\io_1 , \io_2 }_{\io_3,\io_4 }\! &=&\!\Big[U\Big]^{\io_1 , \io_2 }_{\io_3,\io_4 }\!\!+\!\Big[U\frac{1}{1-\chi_0U}\chi_0U\Big]^{\io_1 \io_2}_{\io_3 \io_4}(\kv+\kv') \nonumber \\
&& -\Big[U\frac{1}{1-\chi_0U}\chi_0U  \Big]^{\io_1\io_4}_{\io_3 \io_2}(\kv-\kv') ,
\label{eq:Veff}
\end{eqnarray}
and includes the local bare interactions as well as explicit momentum-dependent effective interactions originating from higher-order processes mediated by the multiorbital susceptibility $\chi_0$. Nesting features of the Fermi surface are important for the pairing structure arising within this mechanism. 

We parametrize the Fermi surface by roughly $3000$ k-points and calculate the superconducting instabilities at the Fermi surface from the linearized gap equation
\begin{eqnarray}
  -\frac{1}{(2\pi)^3}\int_{FS} d \kv_f^\prime \frac{1}{v(\kv_f^\prime)} \Gamma_{l,l'}(\kv_f,\kv_f^\prime)\Delta_{l'}(\kv_f^\prime)=\lambda \Delta_l(\kv_f), \nonumber \\
  \label{eq:LGE}
\end{eqnarray}
where $\Gamma_{l,l'}(\kv_f,\kv_f^\prime)$ denotes the pairing kernel projected to band- and pseudospin-space~\cite{RomerPRL}, and $\kv_f$ denote the wave vectors on the Fermi surface. The Fermi speed is given by $v(\kv_f)$.
The leading superconducting instability is given by the gap function $\Delta_{l}(\kv_f)$ with the largest eigenvalue $\lambda$ and $T_c \propto e^{-1/\lambda}$.
The pseudospin classification is encoded in the subscripts $l,l'=0,x,y,z$ which refer to the components of the ${\bf d({\bf k})}$-vector in pseudospin space~\cite{SigristUeda}. Even parity solutions appear in the $l=0$, pseudospin singlet channel.

\section{Results}

\subsection{Gap structure}
As seen from Fig.~\ref{fig:Cobo1}, the leading and subleading instabilities appear in the $s'$ and $d_{x^2-y^2}$ channels. This result is very robust to parameter variations. In the present case, it is  obtained for the band parametrization stated in Table~\ref{tab:Cobohop}, setting $\lsoc=35$ meV, $t_{12z}=5$ meV and $t_{56z}=3$ meV. The susceptibility entering the pairing kernel in Eq.~(\ref{eq:Veff}) is calculated at $k_{\mathrm B}T=2$ meV. The same two leading gap solutions arise both in the case when pairing is mediated by spin fluctuations, Eq.(\ref{eq:Veff}), and when it is generated directly from onsite attractions (i.e. including only the first momentum-independent part of Eq.(\ref{eq:Veff})) in the case $U'-J<0$. The latter Hund's coupling mechanism requires additionally finite SOC~\cite{Hoshino_2015,vafek,Puetter_2012}. Note, however, since $U'=U-2J$ it makes little sense to go beyond $J/U=0.5$ in the present situation. The detailed structure of the gaps at the Fermi surface is shown in Fig.~\ref{fig:Cobo1}(a,b) and Fig.~\ref{fig:Cobo1}(d,e) for both pairing mechanisms. Note the qualitatively distinct nodal structures supported by the two approaches. First, Fig.~\ref{fig:Cobo1}(a,b) show the solutions of the spin-fluctuation mechanism as formulated in Eq.~(\ref{eq:Veff}). The leading $s'$ solution from Fig.~\ref{fig:Cobo1}(a) features no symmetry-protected nodes, but displays a robust nodal structure which persists for the whole range of explored values of Hund's coupling. The nodal structure on the inner ($\beta$) pocket coincides with the accidental nodes of the subleading $d_{x^2-y^2}$ solution, see Fig.~\ref{fig:Cobo1}(b). This ensures a near-nodal structure on the inner ($\beta$) pocket of a composite order parameter of the form $\Delta_{A_{1g}}+e^{i\phi}\Delta_{B_{1g}}$ (and $\Delta_{A_{1g}}+e^{i\phi}\Delta_{B_{2g}}$), as discussed previously within 2D models for \sruo~\cite{RomerPRL,romer2021}. Second, as opposed to the involved nodal structure supported by the spin fluctuations, which is rooted in the changing orbital content at the Fermi surface~\cite{romer2021}, the Hund's pairing gap structures display only symmetry-dictated nodes. Thus, the leading $A_{1g}$ solution is entirely nodeless as seen from Fig.~\ref{fig:Cobo1}(d), but does exhibit a non-trivial relative sign change between the inner ($\beta$) and two outer ($\alpha,\gamma$) Fermi pockets~\cite{vafek}. Likewise, the  $B_{1g}$ solution features only its symmetry-dictated nodal structure, but also has an additional internal phase structure with a sign change of the simple $d_{x^2-y^2}$ form between the $\alpha,\beta$ and $\gamma$-pockets. As discussed previously, these Hund's pair states correspond to dominant inter-orbital singlet, spin-triplet pairs~\cite{vafek,clepkens_PRB_2021}. The qualitatively distinct gap structures between the two mechanisms arise from significant contributions from the last two terms in  Eq.~(\ref{eq:Veff}). The bare interactions, first term in  Eq.~(\ref{eq:Veff}), contains the attractive Hund's pair hopping term, {\it also} within spin-fluctuation mechanism when $J/U>1/3$, but for the present \sruo~band structure this direct attractive channel is simply overwhelmed by the higher-order processes, even at $J/U=0.4-0.5$. Clearly, this result depends on the combined  significance of interactions and the spin susceptibility supported by the underlying band structure. The competition/cooperation between spin-fluctuation generated solution versus Hund's pairing superconductivity is further explored in Ref.~\onlinecite{Merce}.

As evident from Fig.~\ref{fig:Cobo1}(c), the two-component $E_g$ solution is  suppressed in the spin-fluctuation pairing scenario with several additional 1D even-parity solutions preceding in the hierarchy of superconducting instabilities (not shown in Fig.~\ref{fig:Cobo1})~\cite{Roising2019}. In the Hund's coupling pairing scheme on the other hand, it is the third leading instability, appearing right below the $s$ and $d_{x^2-y^2}$ solutions. The Hund's pairing scenario can support a leading $E_g$ solution, as discussed further below, but only in the unusual case where the onsite SOC is weaker  compared to the k-SOC~\cite{suh2019}. For the current band parametrization with an onsite SOC of $\lsoc=35$ meV and k-SOC terms of magnitude $t_{12z}=5$ meV and $t_{56z}=3$ meV,  the gap structure of one of the two components of the $E_g$ solution is shown in the inset in the bottom of Fig.~\ref{fig:Cobo1}(c). That gap structure is surprisingly similar within both pairing scenarios, i.e. the $E_g$ solutions arising from spin fluctuations do not exhibit additional orbitally driven accidental nodes.

To further compare the two different pairing scenarios, we show in Fig.~\ref{fig:Cobo1}(c) the evolution of pairing eigenvalues as the Hund's coupling $J/U$ increases. In the regime $J/U<1/3$, the local (mean-field) Hund's pairing does not support superconductivity, since no attractive channels are activated. By contrast, spin-fluctuation mediated pairing generates attractive channels for all values of Hund's coupling. We find an increase of the eigenvalues with increasing $J/U$, but no crossover between the different symmetry channels. Note that for all Hund's exchange values $J/U$ in Fig.~\ref{fig:Cobo1}(c), the eigenvalues of the spin-fluctuation pairing significantly dominate over the eigenvalues from the Hund's pairing scenario. In all regimes, we find that odd-parity, pseudo-spin triplet solutions are strongly suppressed. Lastly, we note that within spin-fluctuation-mediated pairing, the strength of the pairing depends on the proximity to the associated magnetic instability at $U_c$, i.e. $U$ compared to the critical interaction strength which in the current case is $U_c\simeq 160$ meV for $J/U=0.2$. Therefore, larger values of Coulomb repulsion cause a rapid increase in the eigenvalue (and hence $T_{\mathrm c}$). This property is only relevant very close to the instability, however, which is not the generic parameter range explored in this paper. 

The gap structures that we find for the spin-fluctuation pairing, Fig.~\ref{fig:Cobo1}(a,b), are reminiscent of the $s'$ and $d_{x^2-y^2}$ gap structures reported previously in purely 2D calculations~\cite{RomerPRL,Romer_strain2020,romer2021}. Compared to the 2D case, there are small differences in the gap structure resulting from the additional interplanar hybridization and k-SOC terms: the gap changes sign on the $\alpha$ pocket and a shift of the accidental node position towards the zone axes is present in the 3D gap structure. However, the detailed gap structure of Fig.~\ref{fig:Cobo1}(a,b) still exhibits a sign change of the gap for segments of the Fermi surface connected by the prominent nesting vector of the $xz$/$yz$ orbitals, namely $\Qv_1\simeq (2\pi/3,2\pi/3,q_z)$.

The gap structures presented in Fig.~\ref{fig:Cobo1}(a,b) are robust to changes in onsite Coulomb repulsion $U$ and Hund's couplings. This leads to the conclusion that, despite the non-trivial 3D structure of the spin susceptibility entering the pairing kernel, the dominant and leading subdominant superconducting instabilities remain $s'$ and $d_{x^2-y^2}$, with a similar gap structure as the ones obtained in the 2D case~\cite{RomerPRL,romer2021}. Due to the demanding numerical procedure, we have not included longer-range interactions in the pairing kernel in the 3D calculations. However, based on the robustness of the $s'$ and $d_{x^2-y^2}$ solutions when expanding the model from 2D to 3D, we expect a similar conclusion to hold when introducing longer-range interactions, in which case spin fluctuations will most likely lead to a near-degeneracy of $s'$ and $d_{xy}$ pairing states~\cite{romer2021}.

\begin{figure}[tb]
    \centering
    \includegraphics[width=\linewidth]{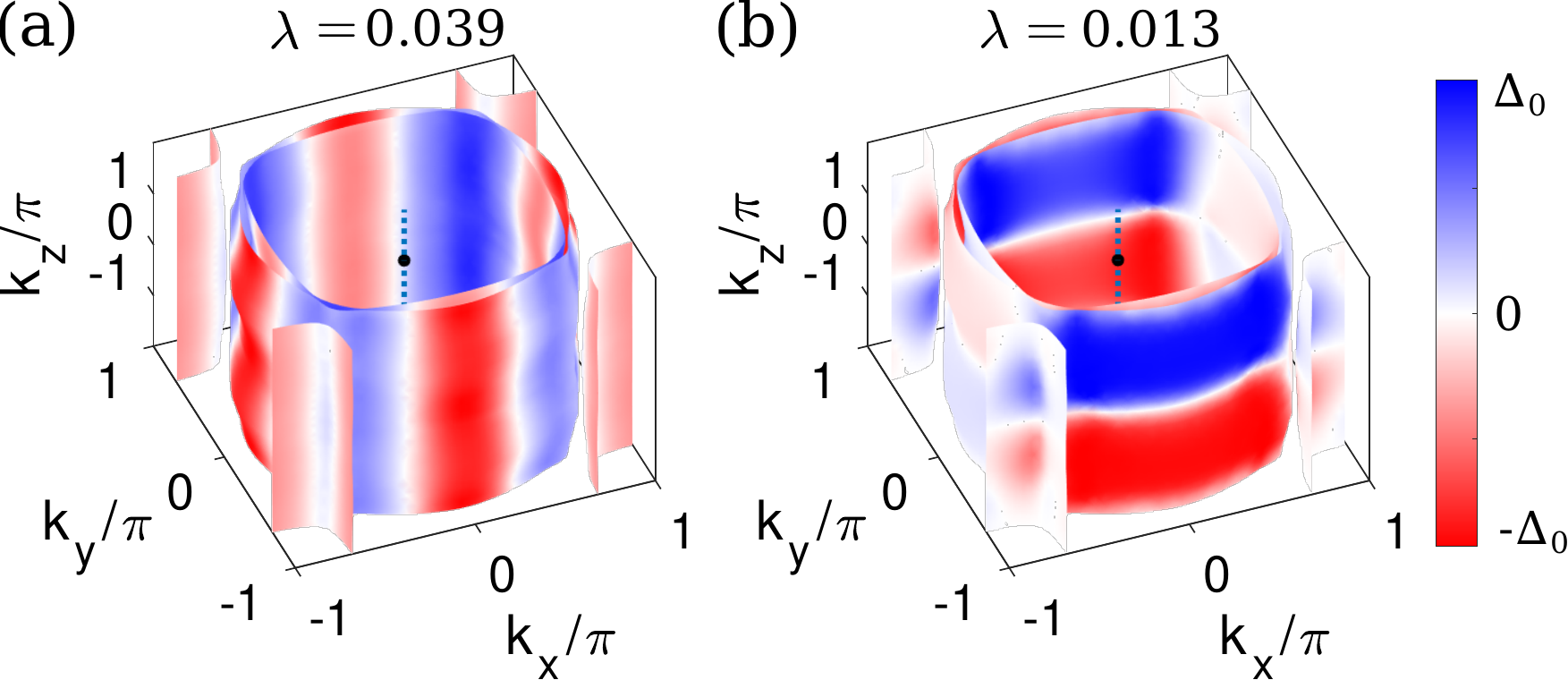}
    \caption{(a) Leading $A_{1g} (s')$ gap structure arising within spin-fluctuation pairing for $U=100$ meV and $J/U=0.5$ for reduced onsite SOC $\lsoc=5$ meV and k-SOC $t_{56z}=4$ meV. (b) Leading $E_{g} (d_{xz}/d_{yz})$ gap structure arising from Hund's coupling mediated pairing (one of the degenerate components shown) for the same parameters as in (a). The corresponding eigenvalue of the linearized gap equation is displayed in each sub-figure.} 
    \label{fig:Cobo9}
\end{figure}

The ratio between onsite SOC and k-SOC is decisive for the appearance of a leading $E_g$ solution within the Hund's pairing mechanism, as explored in Ref.~\onlinecite{suh2019}. In particular, in the regime where $\lsoc$ and  $t_{56z}$ are comparable, the $E_g$ solution is favored when $U'-J<0$. In order to investigate the spin-fluctuation mechanism in the same regime of $\lsoc \simeq t_{56z}$, we show in Fig.~\ref{fig:Cobo9} the leading solutions of the two mechanisms for the case of $\lsoc=5$ meV and $t_{56z}=4$ meV. As expected, the $E_g$ solution is found to be leading within Hund's pairing, in this case for $J/U=0.5$ (Fig.~\ref{fig:Cobo9}(b)). On the other hand, spin fluctuations favor an $s'$ solution for the same set of parameters, i.e. in the strong Hund's coupling regime setting $U=100$ meV despite the large value of $J/U=0.5$. The subleading solution is $d_{x^2-y^2}$ and this continues to be the case also for smaller values of Hund's coupling $J/U=0.25$ (not shown).  Notably, the spin fluctuations yield an eigenvalue which is a factor of three larger than in the Hund's pairing mechanism. Again this value depends on $U$ compared to $U_c$, but it remains a fact that $\lambda$ is significantly larger within the spin-fluctuation approach for a wide range of interaction strengths~\cite{Merce}. Thus, even in the case where parameters are such that $E_g$ is leading within Hund's pairing, the momentum-dependent parts of the pairing kernel Eq.~(\ref{eq:Veff}) overwhelm the gap structure arising purely from the onsite terms.
In addition, we have explored the band structure of Ref.~\onlinecite{suh2019} in the specific case of $\lsoc=20$ meV and $t_{56z}=15$ meV and found that while pure onsite interactions in the regime $J/U=0.5$ supports the $E_g: d_{xz}/d_{yz}$ solution, spin fluctuations give a $d_{x^2-y^2}$ and $s'$ near-degeneracy also for this band parametrization with largely enhanced eigenvalues compared to Hund's pairing.

\subsection{Spin-susceptibility: neutron resonance and NMR Knight shift}

We turn now to a discussion of the properties of the spin susceptibility in the leading superconducting candidate states that currently appear relevant for \sruo. We focus on the question of a neutron resonance and the  temperature $T$ dependence of the NMR Knight shift. 

The low-energy magnetic excitation spectrum at the incommensurate vector $\Qv_1\simeq (2\pi/3,2\pi/3,q_z)$ has been probed in a number of recent neutron scattering studies~\cite{Kunkemoller,Jenni_2021,Iida22}.  Naively it is expected that, whatever its (unconventional) superconducting ground state, 
\sruo ~should exhibit a neutron spin resonance, i.e. an enhancement of spectral weight below the superconducting gap edge $2\Delta$ at momenta that correspond to peaks in the spin susceptibility and which connect regions on the Fermi surface on which the superconducting gap changes sign.  Such a feature has been observed in inelastic neutron scattering on the cuprates, iron-based superconductors, and heavy-fermion systems, among others~\cite{Scalapino2012}. Indeed, in Ref.~\onlinecite{Iida22} a weak resonance-like feature at finite out-of-plane momentum $q_z=0.5$ r.l.u. was reported. However, this finding remains controversial as it has not been reproduced, see e.g. Ref.~\onlinecite{Jenni_2021}, where scattering wave vectors near ${\bf Q}_1$ were carefully probed. Additionally, none of the neutron scattering studies find strong evidence for a pronounced opening of a gap in the spin excitation spectrum.  We believe that this is a key experimental touchstone that may help to identify the structure of the superconducting gap in \sruo.

\begin{figure}[tb]
    \centering
    \includegraphics[width=0.8\linewidth]{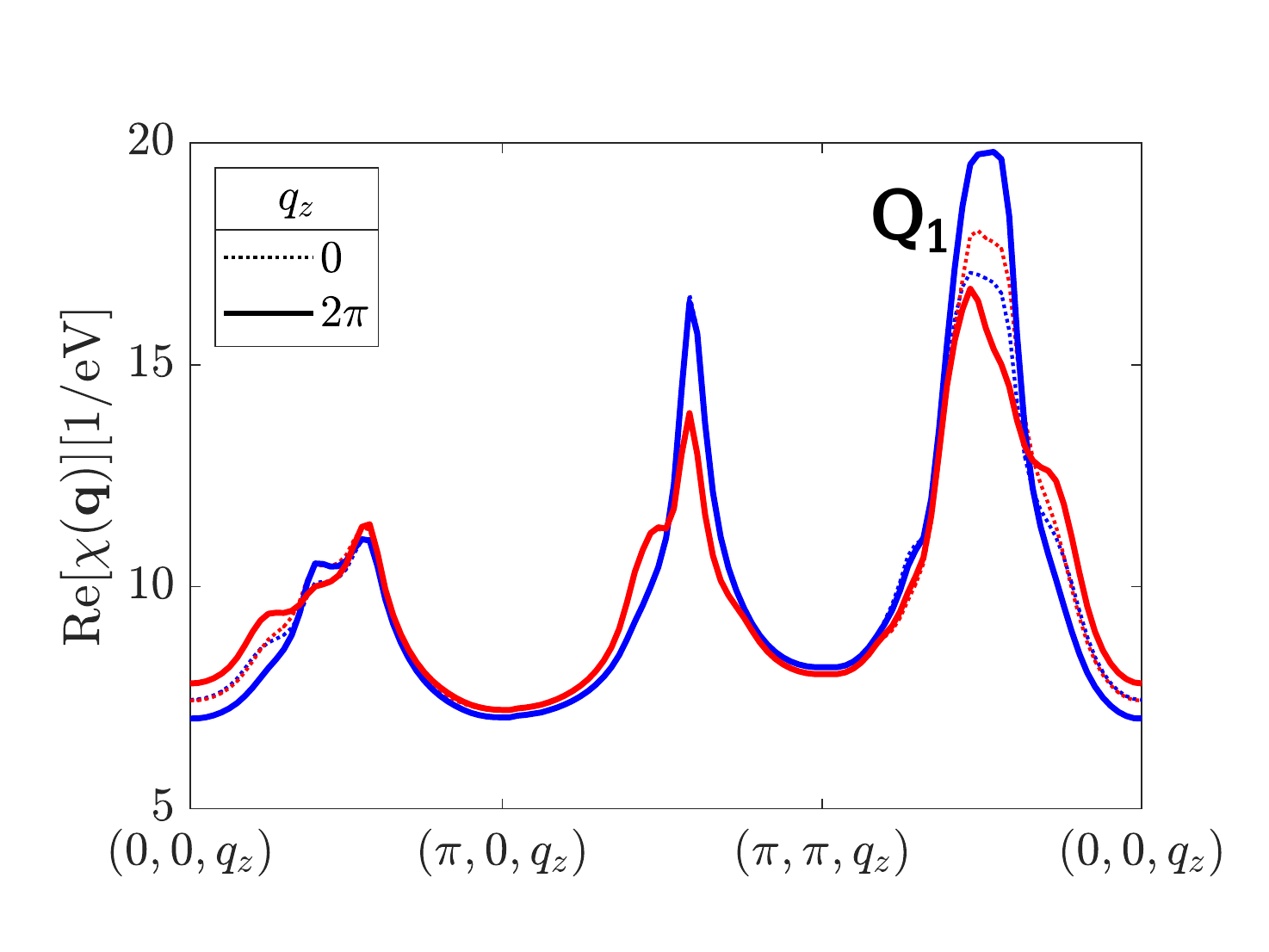}
    \caption{Spin susceptibility of the normal state shown for a momentum path $(0,0,q_z)\to (\pi,0,q_z) \to (\pi,\pi,q_z) \to (0,0,q_z)$ for  $q_z=0$ (dotted lines) and $q_z=2\pi$ (full lines) for $U=100$ meV and $J/U=0.25$. The blue data shows the out-of-plane spin component, $\chi_{zz}$, while red data displays the in-plane spin component, $\chi_{+-}$.} 
    \label{fig:chiRPA}
\end{figure}

In order to address these recent inelastic neutron measurements, we investigate the energy dependence of the spin susceptibility at $\Qv_1$ for two different $q_z$ vectors,  $q_z=0$ and $2\pi$ (note that we have chosen the BZ $[-\pi,\pi]\times[-\pi,\pi]\times[-2\pi,2\pi]$ and therefore $q_z=2\pi$ corresponds to an experimental value of $q_z=0.5$ r.l.u.).
The calculated static $\omega=0$ spin susceptibility in the normal state is shown in Fig.~\ref{fig:chiRPA}, where the effect of SOC is visible in the difference between in-plane and out-of-plane spin components. For earlier theoretical works on the normal state spin susceptibility, we refer to Refs.~\cite{Morr2001,Eremin2002,Tsuchiizu15,Cobo16,RomerPRL,Boehnke_2018,Gingras18,Strand2019,Acharya19}. In the corresponding 2D model, we find a larger out-of-plane spin component at the dominant peak structure at $\Qv_1$~\cite{RomerPRL,romer2021}. Notably, as a result of the new k-SOC couplings, this anisotropy feature is now small and inverted at $q_z=0$, but well-pronounced at $q_z=2\pi$.

We address the possible gap structures proposed by spin-fluctuation mediated pairing, i.e. $s'$, $d_{x^2-y^2}$ (see Fig.~\ref{fig:Cobo1}(a,b)) and the combination $s'+id_{x^2-y^2}$ as well as the Hund's pairing proposed $d_{xz}+id_{yz}$ gap structure, of which one component is shown in Fig.~\ref{fig:Cobo9}(b). We approximate the RPA results for the different gap structures in terms of crystal harmonics. For example, for the $s'$ gap we use
\begin{equation}\label{eq:sPara}
    \Delta(k) = a + b(\cos k_x + \cos k_y) + c(\cos k_x \cos k_y),
\end{equation}
with band-dependent values for the coefficients $a$, $b$ and $c$. Similarly, the $d_{x^2-y^2}$ gap is parametrized in terms of $\cos k_x-\cos k_y$ and $\cos 2k_x-\cos 2k_y$. For these two gap structures, we neglect the $k_z$ dependence, which we find to be negligible in our RPA results. The $E_g$ gap is parametrized in terms of $\sin k_x$, $\sin k_y$, $\sin 2k_x$, $\sin 2k_y$, $\sin k_x\cos k_y$ and $\sin k_y\cos k_x$ crystal harmonics with an overall $\sin k_z/2$ form factor. We then include an additional Gaussian envelope of the form $\mathrm{exp}[-\epsilon_\nu(\kv)/\Omega_0]$ where $\epsilon_\nu(\kv)$ is the eigenenergy of band $\nu$ such that it cuts off the gap for momenta away from the Fermi surface. For all the parametrizations, we use coefficients that lead to a maximum gap size of approximately $\Delta_{\rm max}=10$ meV. Numerical limitations prevent us from reducing the gap size further. This should not be a significant limitation, however, as the normal state density of states in our band structure is featureless over significantly larger energy scales.

\begin{figure}[tb]
    \centering
    \includegraphics[width=\linewidth]{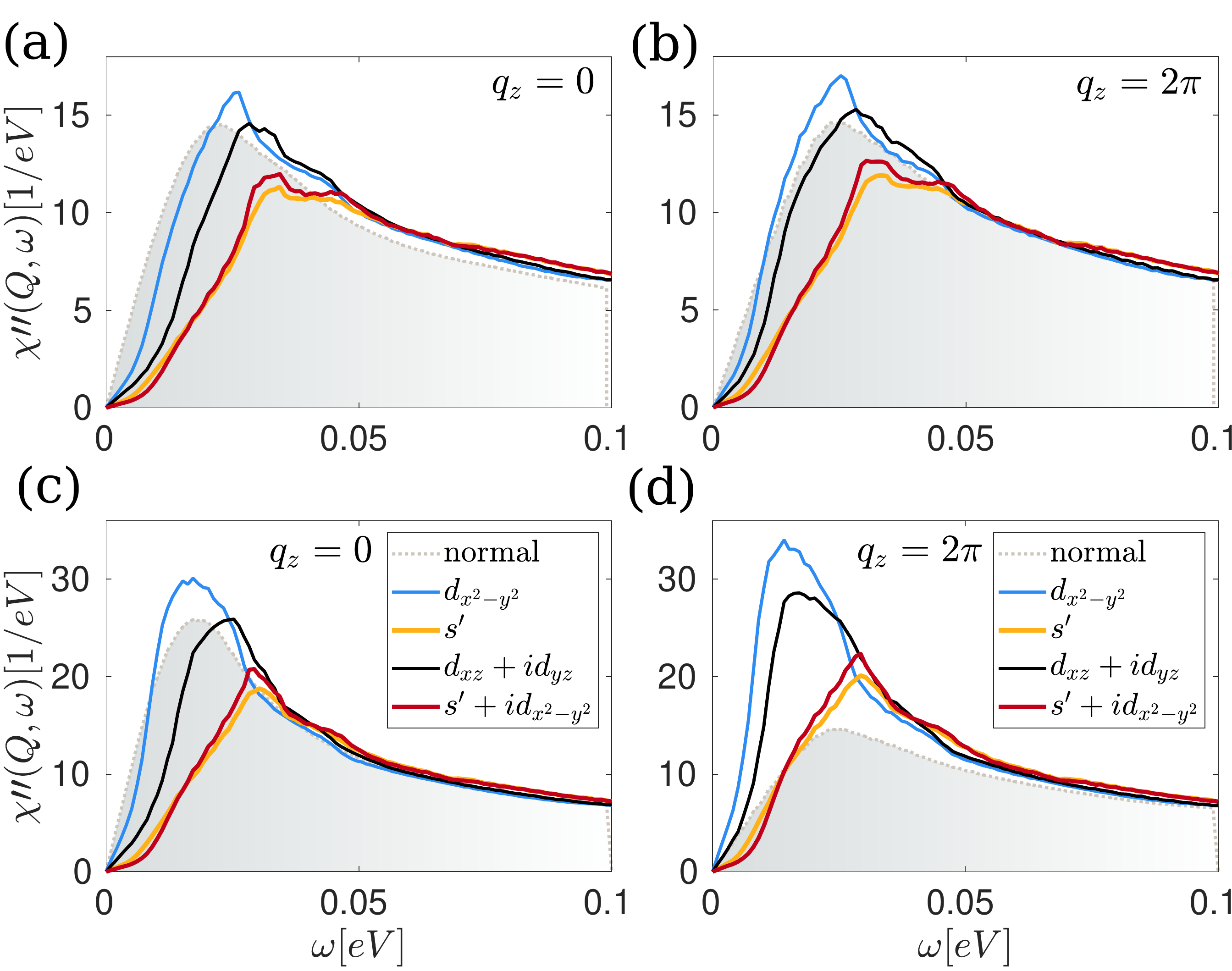}
    \caption{Energy dependence of the imaginary part of the transverse spin susceptibility $\chi''(\Qv_1,\omega)$ in the superconducting state for different gap symmetries $d_{x^2-y^2}$ (blue line), $E_g (d_{xz}+id_{yz})$ (black line), $s'$(orange line) and $s'+id_{x^2-y^2}$ (red line). The normal state spectra are depicted by the shaded area. The bare interaction parameters are (a,b) $U=120$ meV $J/U=0.25$, and (c,d) $U=140$ meV $J/U=0.25$.}
    \label{fig:neutron}
\end{figure}

The calculated low-energy spin susceptibility is shown in Fig.~\ref{fig:neutron} for the different gap structures.  As seen in the depiction in Figs. 1 and 2, the gap magnitudes vary considerably over the Fermi surface sheets. Due to this strong gap anisotropy, there is no clear opening of a well-defined spin-gap $2\Delta$, although there is a suppression of spectral weight at energies below $2\Delta$. At the same time, there is no clear spin resonance at $\Qv_1$ for an interaction strength $U=120$ meV ($J/U=0.25$), see Fig.~\ref{fig:neutron}(a,b). Nevertheless, for larger  interactions ($U=140$ meV) shown in Fig.~\ref{fig:neutron}(c,d), the spectral weight enhancement for the $d_{x^2-y^2}$ gap moves down to lower frequencies below $2\Delta$ and gets significantly stronger for both $q_z=0$ and $q_z=2\pi$, indicating that this is indeed a resonance-like feature. 
Also, the two-component $d_{xz}+id_{yz}$ solution displays a broad resonance feature for larger $U$, but exclusively in the case of $q_z=2\pi$, see Fig.~\ref{fig:neutron}(d). This is expected since for scattering at ${\bf Q}_1$, the gap only changes sign for $2\pi <q_z <4\pi$, see Fig.~\ref{fig:Cobo9}(b). As mentioned above, in the recent study by Iida {\it et al.}, a neutron resonance-like feature was found at this $q_z$ position~\cite{Iida22}. This appears to be in line with our theoretical findings for $d_{xz}+id_{yz}$ superconductivity. 
We stress, however, that the resonance found at $\Qv_1$ for $q_z=0.5$ r.l.u. in Ref.~\onlinecite{Iida22} has not been confirmed by another detailed study of the out-of-plane momentum dependence of the spin fluctuations, see Ref.~\onlinecite{Jenni_2021}. Finally, as seen from Fig.~\ref{fig:neutron}, neither the $s'$ nor the $s'+id_{x^2-y^2}$ solutions exhibit resonance features at any interaction strength or $q_z$ values.

\begin{figure}[t]
    \centering
    \includegraphics[width=0.7\linewidth]{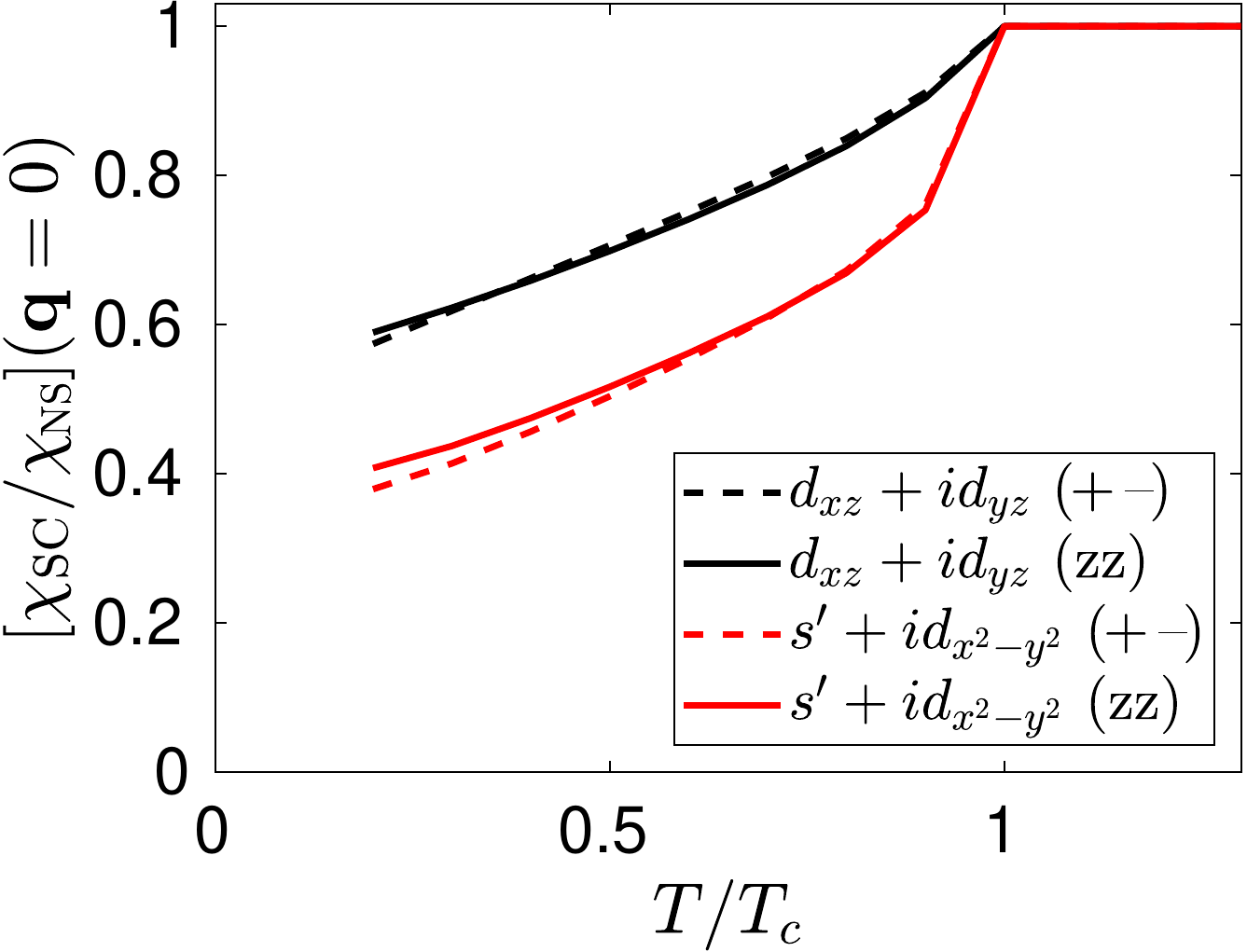}
    \caption{Knight shift calculated as the ratio $\chi_{\rm SC}(\qv=0)/\chi_{\rm NS}(\qv=0)$ at zero energy as a function of temperature $T$ for the two (time-reversal symmetry broken) gap structures $d_{xz}+id_{yz}$ (black curves) and $s'+id_{x^2-y^2}$ (red curves) for $U=140$ meV and $J/U=0.25$. The in-plane (out-of-plane) spin susceptibility is shown by dashed (full) lines. 
    } 
    \label{fig:KS}
\end{figure}

Lastly, we turn to a discussion of the Knight shift which is calculated as the real part of the spin susceptibility at zero momentum, $\chi(\qv=0,\omega=0)$, as a function of $T$ normalized to the normal state value $\chi_{\rm NS}(\qv=0,\omega=0)$. We consider two candidate superconducting gap structures, namely the $d_{xz}+id_{yz}$ state stabilized within Hund's pairing, and the $s'+id_{x^2-y^2}$ phase favored by spin-fluctuation mediated pairing. Both orders break time-reversal symmetry. The $T$-dependence of the gap is introduced explicitly by the BCS form $\Delta(T)=\Delta_0\tanh\left(1.76\sqrt{\frac{T_{\mathrm c}}{T}-1}\right)$, with $\Delta_0=1$ meV at the position of maximum gap magnitude and $T_{\mathrm c}=0.5$ meV. 
Both gap structures display a considerable suppression of the uniform spin susceptibility at the lowest $T$, as shown in Fig.~\ref{fig:KS}. There is only a small difference between in-plane and out-of-plane susceptibility components. As seen from Fig.~\ref{fig:KS}, the decrease is only partial, with the ratio $\nu=\chi(T \ll T_{\mathrm c})/\chi(T > T_{\mathrm c}) \simeq 0.4$ for the $s'+id_{x^2-y^2}$ gap, and $\nu \simeq 0.6$ for the  $d_{xz}+id_{yz}$ gap. Importantly, the recent NMR experiment by Chronister {\it et al.}\cite{Chronister2021} proposes an upper bound of the condensate response to be less than 10 \% of the normal state value. As seen from Fig.~\ref{fig:KS} both time-reversal symmetry broken states violate the "Chronister bound". We have not further explored quantitative match with this here. However, as discussed in Ref. \onlinecite{romer2021} for the case of $s'+id_{xy}$ superconductivity, the magnitude of $\nu$ depends rather sensitively on the amplitude of SOC and interaction parameters $U$ and $J$. Based on these calculations, it appears hard to reconcile a sizable SOC with $\nu < 10$ \%.  

\section{Discussion and conclusions}

Even though the $E_u: p_x/p_y$ spin-triplet order appears ruled out by recent Knight shift measurements~\cite{Pustogow19,Ishida_correct}, several other experiments require certain combinations of two superconducting components~\cite{Xia_2006,ghosh2020thermodynamic,benhabib2020jump,grinenko2021split}. By contrast, specific heat data under uniaxial pressure appears difficult to reconcile with a two-component homogeneous order parameter scenario~\cite{willa2021,Li_2021,Wagner_2021}. Here, we have determined the microscopic nature of superconductivity from spin-fluctuations and Hund's pairing within 3D models for \sruo. We find that $s'$- and $d$-wave even-parity solutions are robust leading superconducting states, similar to the findings from 2D models for this material. This points to the nodal $s'+id$ phase as a prominent candidate for reconciling many experiments. The $E_g: d_{xz}/d_{yz}$ even-parity order is not favored from spin-fluctuations, even inside the Hund's pairing regime at large $J/U$. In addition, we note that the $d+ig$ state suggested on phenomenological grounds~\cite{kivelson2020proposal} also does not appear to be competitive within the current framework.

Hund's pairing as defined in this paper can be viewed as an approximation to spin-fluctuation pairing in the sense that the Hund's pairing kernel is simply the spin-fluctuation kernel without the momentum-dependent terms, see Eq.~(\ref{eq:Veff}). From this perspective, the present finding is that even for $J>U'$ the momentum-dependent terms dominate for an electronic band structure and interaction strengths relevant for \sruo, i.e. they produce significantly larger $T_{\mathrm c}$ and can lead to qualitatively different gap structures. Of course both spin-fluctuation pairing and (mean-field) Hund's pairing are approximate methods to solve the full many-body pairing problem, and it remains to be seen how the superconducting states generated by these respective mechanisms compare to other methods relevant for modelling the superconducting gap structure of \sruo. For recent work along these lines, see e.g. Refs.~\onlinecite{Kaser_2021,Gingras_2022,Gingras2022B}. These works take a large-scale numerical approach to calculation of the post-RPA pairing vertex, predicting different interesting leading even- and odd-frequency pairing states. Thus far they are restricted to high energies, and it is presently unclear how they are related to the low-energy Hamiltonian relevant for a 1K superconductor.

Finally, we computed the spin susceptibility in the superconducting state of leading candidates for \sruo. In contrast to the $s'$ and $s'+id_{x^2-y^2}$ states, it is found that $d_{x^2-y^2}$ and $d_{xz}+id_{yz}$ orders support a sub-2$\Delta$ neutron resonance, which appears in contradiction to the most comprehensive currently-available neutron scattering data~\cite{Jenni_2021}. In terms of the uniform susceptibility, both time-reversal symmetry broken states, $s'+id_{x^2-y^2}$ and $d_{xz}+id_{yz}$, exhibit significant Knight shifts at the lowest $T$, with the former appearing more compatible with recent bounds on the SOC-generated spin-triplet component~\cite{Chronister2021}. Thus, in summary, the mystery of superconducting pairing in \sruo~remains a challenge, but from the perspective of spin-fluctuation mediated pairing the accidental $s'+id_{x^2-y^2}$ (or $s'+id_{xy}$ when including longer-range Coulomb interactions) state currently remains the favored superconducting order, and seems compatible with most experiments.

\section{Acknowledgements}

We thank O. Gingras, M. Roig and H. R\o ising for insightful conversations. A.~T.~R. and B.~M.~A. acknowledge support from the Independent Research Fund Denmark Grant No. 8021-00047B. T.~A.~M. was supported by the U.S. Department of Energy, Office of Basic Energy Sciences, Materials Sciences and Engineering Division. P.~J.~H. was supported by the U.S. Department of Energy under Grant No. DE-FG02-05ER46236. 
 \bibliography{bibliography_SRO3D}
\clearpage
\appendix
\twocolumngrid

\end{document}